\documentstyle[12pt]{article}
\begin{document}
\newcommand{\noi}{\noindent}
\newcommand{\eq}{\begin{equation}}
\newcommand{\en}{\end{equation}}
\newcommand{\eqa}{\begin{eqnarray}}
\newcommand{\ena}{\end{eqnarray}}
\newcommand{\smgr}{\stackrel{\textstyle <}{>}}
\newcommand{\grsm}{\stackrel{\textstyle >}{<}}
\newcommand{\aleq}{\mbox{}_{\textstyle \sim}^{\textstyle < }}
\newcommand{\ageq}{\mbox{}_{\textstyle \sim}^{\textstyle > }}
\newcommand{\ra}{\rightarrow}
\newcommand{\lra}{\longrightarrow}
\newcommand{\bt}{\beta}
\newcommand{\bre}{\hfill\break}
\renewcommand{\thefootnote}{\fnsymbol{footnote}}
\setcounter{footnote}{1}
\noindent July 1994   \hfill HU Berlin--IEP--94/11
\par
\vspace*{2cm}
\begin{center}
{\LARGE Compact Lattice QED with Staggered Fermions and
Chiral Symmetry Breaking}
\footnote{Work supported by the Deutsche
Forschungsgemeinschaft under research grant Mu 932/1-2} \\
\par
\vspace*{2cm}
{\large
A.~Hoferichter, $\mbox{}$
V.K.~Mitrjushkin $\mbox{}$
\footnote{Permanent adress: Joint Institute for Nuclear Research,
Dubna, Russia},
M.~M\"uller--Preussker $\mbox{}$
}
\par
\vspace*{0.5cm}
{\normalsize
$\mbox{}$ {\em Humboldt--Universit\"at, Institut f\"ur Physik,
10099 Berlin, Germany}}  \\
\par
\vspace*{2cm}
{\bf Abstract}
\end{center}
\par
\bigskip
Different formulations of the $4d$
compact lattice QED with staggered fermions  (standard Wilson and
modified by suppression of lattice
artifacts) are investigated by
Monte Carlo simulations within the quenched approximation.
We show that after suppressing lattice artifacts
the system undergoes a phase transition from the Coulomb phase into a
presumably weakly chirally broken phase only at (unphysical) negative
$\beta$--values.
\par
\vspace*{2cm}
\noindent
{\bf i)}
The lattice formulation of any field theory is not unique.
On a physical ground one has to decide which version is
realized in nature, if different lattice versions do not
belong to the same universality class.
\par
At present lattice QED is mostly discussed within a non--compact
realization of the gauge part in the lattice action \cite{qed1,qed2}.
However, considering QED as arising from a subgroup of a non--abelian
(e.g. grand unified) gauge theory we are led to a {\it compact}
description of the local gauge symmetry.
\par
The aim of this note is to shed some light on the
mechanism of spontaneous chiral symmetry breaking (SCSB) in
lattice gauge theories.
Compact lattice QED with staggered fermions
serves very well for this purpose. It is comparatively
simple, and in the limit of vanishing fermion mass the classical action is
chirally symmetric. In case of the standard Wilson gauge action the
quantized theory
undergoes a phase transition, such that in the zero-mass limit  the
chiral condensate $~ \langle {\bar \chi} \chi \rangle~$
becomes zero within the Coulomb phase (i.e. at $~\beta~\equiv~1/e_{bare}^2
>\beta_0~\simeq~1.0~$), and
non--zero in the confinement phase ($~\beta < \beta_0~$)
\cite{kog1}.
\par
The confinement phase of compact QED does not seem to be realized in nature.
It is related to the existence and condensation of De Grand--Toussaint
monopoles~\cite{gt} which we can understand to be artifacts of the realization
of the lattice discretization.
\par
In case of the pure gauge theory
it has been shown \cite{bss,bmm,bmmp} that a complete suppression
of these monopoles removes the phase transition at $~\beta_0~$.
One is left with a unique Coulomb phase (at least at positive $~\beta$'s)
\cite{bmm}.
A further modification of the theory by suppressing
additionally negative plaquettes
was shown not to change the spectrum of gauge degrees of freedom but
to improve the overlap of the plaquette operator with the
lowest state, at least, in the case of the photon quantum numbers \cite{bmm}.
\par
A {\it comparative} study of the chiral properties
of the standard and of the modified theories
with compact gauge fields coupled to
staggered fermions is the aim of this investigation.
Within the quenched approximation we
are going to discuss the phase structure of these theories. We will show that
the  modified theories undergo phase transitions at  negative
(i. e. unphysical) $~\beta$--values only.
\par
\vspace*{1cm}
\noindent
{\bf ii)}
We consider compact $~U(1)~$ gauge fields
coupled to staggered fermions defined
on a lattice of the size $~L_s^3 \cdot L_t~$  (different time and spatial
extensions are only introduced for the computation of correlators).
The gauge fields are treated with the Wilson action modified by suppressing
lattice artifacts, such that the gauge action in general reads as follows
\eq
   S_{G}(U_{l})
   =  \beta \cdot \sum_{P} \left( 1 - \cos \theta_{P}\right)
   + \lambda_{K} \cdot \sum_{c} |K_{c}|
   + \lambda_{P} \cdot \sum_{P} \bigl( 1 - {\rm sign}(\cos \theta_{P}) \bigr)~,
                                              \label{amod}
\en
\noi
where
  $~U_l \equiv U_{x \mu} = \exp (i \theta_{x \mu} ),
   \quad \theta_{x \mu} \in (-\pi, \pi] ~$
are the field variables defined on the links $l \equiv (x,\mu)~$.
The plaquette angles are
  $~\theta_{P} \equiv \theta_{x;\, \mu \nu}
   = \theta_{x;\, \mu} + \theta_{x + \hat{\mu};\, \nu}
   - \theta_{x + \hat{\nu};\, \mu} - \theta_{x;\, \nu} ~$.
\par
The second and the third term in the action have been added in order 
to suppress monopoles and negative plaquette values, respectively,
as lattice artifacts. The monopole currents $~K_c~$ are defined as follows.
The plaquette angle can be splitted
\eq
   \theta_{P} = {\bar \theta}_{P} + 2 \pi n_{P},
   \quad -\pi~<~{\bar \theta}_{P}~ \le ~\pi,
   \quad n_{P}~=~0, \pm 1, \pm 2,
\en
where $~{\bar \theta}_{P}~$ describes the (gauge-invariant)
'electromagnetic' flux
through the plaquette and $~n_{P}~$ is the number of 'Dirac' strings passing
through it. The net number of Dirac strings passing through the surface of
an elementary $~3d~$ cube $~c_{x,\mu}~$ and taken with the correct
mutual orientation determines the monopole charge within this cube,
i.e., a monopole current along the corresponding dual link labeled by
$~(x,\mu)~$
\eq
   K_{c} \equiv K_{x,\mu}~=~ \frac{1}{2\pi} \sum_{P} {\bar \theta}_{P}
   ~=~ \frac{1}{2\pi} \sum_{P} \bigl( \theta_{P} - 2\pi n_{P} \bigr)
   ~=~ -\sum_{P} n_{P}~.
\en
\par
In the action (\ref{amod}) the parameters $~\lambda_{K}~$ and
$~\lambda_{P}~$ play the role
of chemical potentials for monopoles and negative plaquettes, respectively.
In the present investigations we have chosen the following cases
$~(\lambda_{K},\lambda_{P})~=~(0,0)~$ -- standard Wilson action
WA, $~(\infty,0)~$ -- modified action MA1 with total monopole
suppression, and $~(\infty,\infty)~$ -- modified action MA2 with
simultaneous total monopole and negative plaquette suppression.
In practice we realize this total suppression via Kronecker $~\delta$'s in the
measure of the functional integral. They are easily locally taken into account
in the heat bath or Metropolis Monte Carlo updating procedure.
\par
The staggered fermion part of the action looks as follows.
\eq
   S_{F}(U_{l}, {\bar \chi}, \chi)
   ~=~- {\bar \chi}(x) \bigl( {\cal M} + m \bigr)_{xy} \chi(y)
\en
\eq
   {\cal M}_{xy}
   ~=~ - \frac{1}{2} \sum_{\mu}
     \eta_{\mu}(x)  \bigl[ U_{x\mu} \delta_{y,x+{\hat \mu}}
   - U^{\dagger}_{y\mu} \delta_{y,x-{\hat \mu}} \bigr]
\en
where 
$~\eta_{\mu}(x)~=~(-1)^{x_1+x_2+ \dots +x_{\mu-1}}, \quad \eta_1(x)~=~1~$,
and $~\chi~$, $~{\bar \chi}~$ represent one-component Grassmanian variables.
\par
Gauge invariance requires the gauge fields to couple to fermions
via the phase factors 
$~U_{x\mu}~$, i.e., 
in a compact
way. This holds independently of the compact or non--compact
realization of the gauge field action $~S_{G}~$. Therefore, one should keep
in mind that lattice QED within the standard compact and the non--compact case
agree at strong coupling $~\beta~=~0~$. On the contrary, the constraints due
to the modification terms in MA1 and MA2 change the strong coupling behaviour.
Therefore, from the beginning at strong coupling we cannot a priori expect
qualitative agreement with non--compact QED.
\par
Here, we treat the fermions within the quenched approximation. We expect
that the chiral properties of the full theory will be qualitatively correctly
described.
\par
First of all we are going to discuss bulk properties, i.e., we measure
the average plaquette action
$~E_{P}~\equiv~\langle 1- \cos \theta_{P}\rangle~$ as well as
the chiral condensate
$~\langle {\bar \chi} \chi \rangle~=
~\frac{1}{V}~\langle ~\mbox{Tr} \bigl( {\cal M }+ m \bigr)^{-1} \rangle~$.
The latter is computed applying the conjugate
gradient algorithm combined with the noisy estimator method.
\par
Moreover, in order to distinguish the Coulomb phase for the modified
actions even in the negative $~\beta$--range we calculated the
$J^{PC}~=~1^{+-}~$ plaquette--plaquette correlator, 
which for non--vanishing momenta
$~p_{i}~=~2 \pi k_{i} / L_s~$ with $~ k_{i}~= \pm 1, \pm 2, \ldots~$
contains the photon--state ($~1^{--}~$) as the leading contribution.
The photon correlator $~\Gamma(\tau)~$ is 
\eq
\Gamma(\tau)
 = \langle \, \Phi_{\vec{p}}^{\ast}(t+\tau) \cdot \Phi_{\vec{p}}(t)
\, \rangle_c \;,
                                         \label{gamma1}
\en
\noindent 
where the operators 
$~\Phi_{{\vec p}}(t)~$ 
are defined as
Fourier transforms of $~\sin \theta_P~$ \cite{bmm,bp}.
This correlator is expected to behave as
\eq
\frac{\Gamma(\tau)}{\Gamma(0)}~=~A \cdot
\Bigl[~ \exp \Bigl(- \tau \cdot E(\vec{p}) \Bigr) +
\exp \Bigl(- (L_{t}-\tau) \cdot E(\vec{p}) \Bigr)~\Bigr] + \ldots ~,
\qquad 0\leq A \leq 1
                                          \label{gamma3}
\en
with the energy $~E(\vec{p})~$ in units of the lattice spacing given 
by the lattice dispersion relation
\eq
 \sinh^{2} \frac{E}{2} = \sum_{i=1}^{3} \sin^{2} \frac{p_{i}}{2} ~
                                                \label{ldr}
\en
for a zero--mass excitation.

\par
\vspace*{1cm}
\noindent
{\bf iii)}
First of all we have studied the phase structure of the two modified
theories (MA1 and MA2) in comparison with the standard theory (WA).
The simplest way is to consider 
the average plaquette $~E_{P}~$ as
a function of $~\beta~$. We extend the range to be investigated to negative
(i.e., a priori unphysical) $~\beta~$ values.
\par
For the standard theory the plaquette action shows
two transition points at $~\beta~\simeq~\pm 1.0~$, related to the
symmetry $~\beta ~\rightarrow~ -\beta~$ and
$~U_{x \mu}~\rightarrow~\eta_{\mu}(x) \cdot ~U_{x \mu}~$
(full circles in Fig.1a).
In the same figure we have plotted (open squares) the result for
the modified action MA1. Obviously the transition at positive $~\beta~$
is removed leaving us only with a rather strong discontinuity at
$~\beta~\simeq~-0.7~$. For the further modification by suppressing
additionally negative plaquettes (MA2) this discontinuity is shifted
even further down to a $~\beta_{c}~\simeq~-1.68~$
(full squares in Fig.1a).
There we have seen a clear bi--stable behaviour 
indicating the
existence of a first order phase transition (Fig.1b).
\par
The interesting lesson from these calculations is that
in the strong coupling region (as well as
at negative $~\beta$'s) there is a very strong influence of
the lattice artifacts considered.
\par
We can ask, whether the negative $~\beta$--range
$~\beta_{c}~<~\beta~\le~0~$ can be interpreted
as an (analytical) continuation of the Coulomb phase at positive $~\beta~$.
In order to answer this question we computed the photon propagator as explained
before. The results are shown in Fig.2. For the case MA2 we have plotted
$~\Gamma(\tau)~$ for $~\beta~=~0.9~$ and $~\beta~=~-1.1~$, for comparison, at
lowest non--vanishing momentum. The theoretically expected zero--mass curve
is also shown (broken line). All the data points are in perfect
agreement with a zero--mass photon.
\par
In Fig.3 we have plotted the chiral condensate
$~\langle {\bar \chi} \chi \rangle~$
for fixed fermion mass $~m~=~0.04~$ (in units of the lattice spacing)
as a function of $~\beta~$ for the standard action WA and for the modified
one MA2. Whereas for WA $~\langle {\bar \chi} \chi \rangle~$
shows clearly chiral
symmetry breaking at $~\beta~\simeq~1.0~$ (as well as a sign of the
symmetric transition at $~\beta~\simeq~-1.0~$)
it stays small and approximately
constant for the modified action down to $~\beta_{c}~\simeq~-1.68~$.
Thus, it does not differ from the behaviour at $~\beta~>~1.0~$, i.e., in the
Coulomb phase. However, we observed fluctuations of the chiral order
parameter becoming stronger the nearer we came to the phase transition point
$~\beta_{c}~$. At $~\beta~<~\beta_{c}~$ the fluctuations are very strong
making it hard to measure $~\langle {\bar \chi} \chi \rangle~$ in 
the zero--mass limit.
The rise of the chiral condensate in this range, nevertheless,
could indicate
the possibility of a weak spontaneous chiral symmetry breaking.
\par
At several $~\beta~$ values we have studied the mass dependence
of the chiral condensate in the range $~m~=~0.01~\dots~0.05~$ for different
lattice sizes up to $~16^4~$ (see Table).
As an example in Fig.4 we have plotted the mass
dependence for $~\beta~=~0.~$
The finite size effects turn out to be small. The extrapolation down to
$~m~=~0~$ provides a chiral condensate value compatible with zero. An
analogous behaviour has been seen at $~\beta~=~0.5,~1.1$ and~$-1.1~$.
\par
\vspace*{1cm}
\noindent
{\bf iv)}
Our main conclusion is the following. At real bare
coupling, i.e., for positive $~\beta~=~1/e_{bare}^2~$
the modified theory MA2 (at least in quenched approximation) is
fully in the Coulomb phase, and there is no indication for a
phase transition and for spontaneous chiral symmetry breaking.
Therefore, we conclude that SCSB in the standard theory
is due to lattice artifacts.
This is similar to the situation of compact QED with
Wilson fermions \cite{hmmn,hmmns}.
\par
One can try an analytical continuation to negative $~\beta~$ values.
We have seen the Coulomb phase to be extended down to a certain
value $~\beta_c~<~0~$ depending on the concrete realization of the modified
gauge action (called MA1 or MA2). There, we observed a phase transition
presumably of first order. The phase beyond $~\beta_c~$ seems to
exhibit weak chiral symmetry breaking. But, more thorough investigations
with very high statistics are required before drawing any conclusions
in  this respect.
\par
The question  arises if the inclusion of dynamical fermions could change
the picture. However, presumably one should not expect qualitative changes
as we know from studies of both compact and non--compact QED cases.
\par
\vspace*{1cm}
\noi
{\large \bf Acknowledgements}
\par\bigskip\noi
We gratefully acknowledge that the DESY/HLRZ group provided us with the
staggered fermion conjugate gradient algorithm used throughout this work.

\par
\newpage
\noi
{\large \bf Table}
\par\bigskip\noi
The chiral condensate $~ \langle {\bar \chi} \chi \rangle~$ for MA2 at 
different values of $~\beta,~m~$ and for different lattice sizes.
The given errors are purely statistical.
\par\bigskip
\noi
\mbox{}
\vbox{\tabskip=0pt \offinterlineskip
\def\tablerule{\noalign{\hrule width13.82cm}}
\halign to125pt {\strut#&
  \vrule# \tabskip=0.7em plus2em&
  \hfil#\hfil& \vrule#&
  \hfil#\hfil& \vrule#&
  \hfil#\hfil& \vrule#&
  \hfil#\hfil& \vrule#&
  \hfil#\hfil& \vrule#&
  \hfil#\hfil& \vrule#&
  \hfil#& \vrule#\tabskip=0pt\cr\tablerule
&&\omit\hidewidth $\bt$\hidewidth&&
 \omit\hidewidth $L$\hidewidth&&
 \omit\hidewidth $m=0.01$\hidewidth&&
 \omit\hidewidth $m=0.02$\hidewidth&&
 \omit\hidewidth $m=0.03$\hidewidth&&
 \omit\hidewidth $m=0.04$\hidewidth&&
 \omit\hidewidth $m=0.05$\hidewidth&\cr\tablerule\cr\tablerule
&& 0.0&& 8&&.0099(6)&&.0198(8)&&.0299(7)&&.0378(6)&&.04601(8)&\cr\tablerule
&& 0.0&&12&&.00956(5)&&.0213(5)&&.0286(2)&&.0380(4)&&.0469(2)&\cr\tablerule
&& 0.0&&16&&.00934(2)&&.01923(9)&&.0288(1)&&.03748(7)&&.04644(6)&
\cr\tablerule\cr\tablerule
&& 0.5&& 8&&.00854(2)&&.01705(4)&&.02554(3)&&.03432(8)&&.04264(6)&
\cr\tablerule
&& 0.5&&12&&.0090(1)&&.01739(9)&&.0277(4)&&.0360(3)&&.0430(2)&\cr\tablerule
&& 0.5&&16&&.00871(4)&&.0181(2)&&.0262(1)&&.0345(1)&&.0450(3)&
\cr\tablerule\cr\tablerule
&& 1.1&& 8&&.00784(4)&&.0159(1)&&.0239(2)&&.0328(3)&&.0404(6)&\cr\tablerule
&& 1.1&&12&&.008124(6)&&.0169(3)&&.0257(4)&&.0316(3)&&.0394(2)&\cr\tablerule
&& 1.1&&16&&.0085(2)&&.0171(2)&&.0245(2)&&.0331(2)&&.0417(2)&
\cr\tablerule\cr\tablerule
&&-1.1&&12&&.01260(4)&&.02376(3)&&.03595(4)&&.04651(4)&&.05967(6)&
\cr\tablerule
&&-1.1&&16&&.01202(1)&&.02384(2)&&.03556(4)&&.04729(4)&&.05879(4)&
\cr\tablerule\cr\tablerule
&&-2.0&&12&&.0343(2)&&--&&.0852(3)&&--&&.1252(2)&\cr\tablerule
&&-2.0&&16&&.02822(4)&&.0585(1)&&.0807(1)&&--&&.1236(1)&\cr\tablerule
\noalign{\smallskip}
\hfil
\cr}}
\par
\vspace*{2cm}
\noi
{\large \bf Figure Captions}
\par\bigskip\noi
{\bf Fig. 1a:} Plaquette energy $~E_{P}~$ as a function of $~\beta~$
for standard
Wilson action WA and modified actions MA1, MA2. Lattice sizes are $~6^4~$
for WA and MA2 and $~4^4~$ for MA1.
\par\noi
{\bf Fig.1b:} Time history for the plaquette energy $~E_{P}~$ at the 
transition \bre
point $\beta_{c}=~-1.68$ for the modified action MA2. The lattice size
is $~6^4~$.
\par\noi
{\bf Fig.2:} The photon correlator $~\Gamma(\tau)~$ measured at
$~\beta~=~0.9~$  and $~-1.1~$ with lowest non--vanishing momentum
on a $~12 \cdot 6^3~$-- lattice for MA2.
\par\noi
{\bf Fig.3:} Chiral order parameter $~\langle {\bar \chi} \chi \rangle~$ 
as a function of $~\beta~$ for fixed fermion 
mass $~m~=~0.04~$ for WA and MA2, respectively.
The lattice size is $~6^4~$.
\par\noi
{\bf Fig.4:} Mass dependence of 
$~\langle {\bar \chi} \chi \rangle~$ for
the modified action MA2 at $~\beta~=~0.~$ and different 
lattice sizes ($~8^4, 12^4, 16^4~$).

\end{document}